# Reliable and Broad-range Layer Identification of Au-assisted Exfoliated Large Area MoS$_2$ and WS$_2$ Using Reflection Spectroscopic Fingerprints

Bo Zou[1], Yu Zhou[1], Yan Zhou[2], Yanyan Wu[1], Yang He[1], Xiaonan Wang[1], Jinfeng Yang[1], Lianghui Zhang[1], Yuxiang Chen[1], Shi Zhou[3], Huaixin Guo[4], Huarui Sun[1,*]

[1] *School of Science and Ministry of Industry and Information Technology Key Laboratory of Micro-Nano Optoelectronic Information System*

*Harbin Institute of Technology Shenzhen, 518055, P. R. China*

*Collaborative Innovation Center of Extreme Optics, Shanxi University, Taiyuan, Shanxi 030006, People's Republic of China*

[2] *State Key Laboratory of Superlattices and Microstructures, Institute of Semiconductors, Chinese Academy of Sciences*

*Beijing 100083, P. R. China*

[3] University of Science and Technology of China, Hefei 230026, P. R. China

[4] Science and Technology on Monolithic Integrated Circuits and Modules Laboratory, Nanjing Electronic Devices Institute, Nanjing 210016, P.R. China

[*]Corresponding author. Email address: huarui.sun@hit.edu.cn

**Abstract:** The emerging Au-assisted exfoliation technique provides a wealth of large-area and high-quality ultrathin two-dimensional (2D) materials compared with traditional tape-based exfoliation. Fast, damage-free, and reliable determination of the layer number of such 2D films is essential to study layer-dependent physics and promote device applications. Here, an optical method has been developed for simple, high throughput, and accurate determination of the layer number for Au-assisted exfoliated MoS2 and WS2 films in a broad thickness range. The method is based on quantitative analysis of layer-dependent white light reflection spectra, revealing that the reflection peak intensity can be used as a clear indicator for determining the layer number. The simple yet robust method will facilitate the fundamental study on layer-dependent optical, electrical, and thermal properties and device applications of 2D materials. The technique can also be readily combined with photoluminescence and Raman spectroscopies to study other layer-dependent physical properties of 2D materials.

## 1. Introduction

Two dimensional (2D) transition metal dichalcogenides (TMDCs), such as MoS$_2$,

WS$_2$, MoSe$_2$, and WSe$_2$, have attracted much attention due to their unique optical, electrical, thermal, mechanical and spin properties[1-11], which are highly dependent on the layer number of the van der Waals structures. Mechanical exfoliation has been the most commonly used method to obtain high-quality single-crystalline monolayer and few-layer 2D materials, although the exfoliated sample size is quite limited.[12] Very recently, a novel Au-assisted mechanical exfoliation technique has emerged that can be applied to dozens of 2D crystals.[13-18] Using this method, not only high quality millimeter-size or even centimeter-size 2D TMDCs monolayers can be obtained, but separated large-area few-layer nanoflakes, which are needed in hybrid functional devices in photodetection and photocatalysis,[19-23] can also be acquired. Within these large-area nanoflakes, there usually exist abundant regions of different thicknesses varying by monolayer and easy to distinguish, which is otherwise difficult to achieve in tape-exfoliated samples. This type of nanoflakes provides an excellent platform for the investigation of layer-dependent physics and for the demonstration of proof-of-concepts devices, for which reason the Au-assisted exfoliation technique is stimulating the growing interest of the 2D community. For either purpose, it is essential to determine the thicknesses of as exfoliated plentiful nanofilms rapidly, reliably and non-destructively.

So far, a variety of thickness identification techniques of 2D nanofilms have been exploited, including atomic force microscopy (AFM),[12] scanning electron microscopy (SEM),[24] transmission electron microscopy (TEM),[25] scanning tunnelling microscopy (STM),[26] Raman spectroscopy,[27-29] photoluminescence (PL) spectroscopy,[5] and optical microscopy (OM).[24, 30-40] Among these techniques, AFM is most widely used to directly measure the thickness of 2D nanoflakes with small size, but it is time-consuming and low-throughput for large area films. Moreover, AFM is a contact method that can potentially induce structural defects[41] and the measured results might be affected by the absorbed water layer. Electron microscopy (SEM, TEM, STM) is not only time-consuming and costly, but may also introduce pollution or damage due to electron beam induced deposition or atomic displacement.[42-43] Raman and PL spectroscopies can identify TMDCs nanofilms quickly; however, their applications are normally rather limited to distinguishing very thin samples between 1-6L[5, 27] because these two spectroscopy techniques are based on the layer dependence of the lattice structure or electronic structure, which are close to bulk for thick samples. Moreover, PL spectroscopy generally fails for Au-assisted exfoliated TMDCs nanofilms due to PL quenching induced by additional nonradiative recombination paths.[44] In view of all this, a high-throughput, high-accuracy methodology for thickness determination needs to be developed for large area films with various thicknesses.

Intriguingly, Optical microscopy (OM) is a simple, reliable, non-contact and non-destructive technique that enables rapid and high-throughput characterization of large area 2D nanofilms. The OM method is mainly divided into two categories:[45] one is based on the apparent color of the samples[24, 30-33] and the other is based on optical contrast.[34-40] The color-based OM approach requires complex calculations and depends sensitively on the light source and the substrate.[33] The optical contrast based approach uses either the optical contrast between the 2D material and the substrate measured from the images' RGB channels[38-40] or the wavelength dependent optical contrast spectra.[34-37] The RGB method faces challenges in determining precise thickness variations, especially in regions with infinitesimal contrast variation due to the low spectral resolution provided by the RGB filters.[46] Contrarily, with the help of the grating, the spectroscopy technique can provide complete information on the spectral dependence of optical contrast in the visible light band. It is worth noting that in addition to determining the thickness, the optical contrast spectra can also be used to study the optical properties of 2D materials, such as the band gap, excitonic effects, dielectric function and absorption spectra.[47-50] It can further be extended to detecting strain, interlayer charge transfer, stacking order, and twisted angle of 2D materials as these factors can affect the electronic band structure.[51-54]

In this work, we introduce a novel method that can successfully combine the advantages of existing optical techniques to accurately determine the thickness of TMDCs nanofilms. By choosing the highly controllable reflection spectra measurement mode of the Raman spectrometer, no additional instruments are needed and comprehensive layer-dependent white light reflection spectra with more precise and richer information can be obtained. With high-resolution and wide-range spectra, a series of exciton-induced reflection peaks from $MoS_2$ and $WS_2$ nanofilms were observed, whose both peak positions and intensities have obvious thickness dependence. From the quantitative analysis of intensities of both reflection peaks from samples and emission peaks from the LED light source, we have proposed a reliable and broad-range thickness determination method of TMDCs films that simply uses the reflection peak intensity without complicated data processing. Fresnel theory based theoretical calculations of layer-dependent reflectance, which show good agreement with our experimental results, revealed that the physical origin of this method is the classical interference effect. The reliability, generalizability and advantages of the method have been verified by AFM and Raman measurements.

**2. Results and discussion**

The optical microscope images of Au-assisted mechanical exfoliated large area

MoS$_2$ nanofilms with various thickness regions (1-15L, confirmed by AFM measurements in Figure S1) are shown in **Figure 1**a-c. The white light reflection spectra (WLRS) in the range of 400-950 nm were measured by a Raman spectrometer (see the Experimental Section for details). The white light focal spot radius was determined to be 1.72 ± 0.03 μm by the knife edge method[9, 50, 55] (Figure S2). Figure 1d shows a WLRS mapping on a MoS$_2$ nanofilm with 1-5L regions; the clear boundaries between different regions and the uniform color gamut of the same thickness region indicate the high spatial resolution and good repeatability of this method, respectively. The layer-dependent WLRS measured from MoS$_2$ nanofilms as well as from the Au/Ti/SiO$_2$/Si substrate (see the Experimental Section for details, abbreviated as the Au substrate in the following) are shown in Figure 1e (1-10L). The WLRS of 1-35L MoS$_2$ in the Supporting Information reveals that each region with a different layer number has a unique reflection spectral fingerprint. This is unlike the color-based method that may generate indistinguishable colors for regions with minimal layer number difference or even periodic color changes upon certain thickness variations.[30, 56] In the reflection spectra of the bare Au substrate, the three peaks located at 450, 605, and 900 nm correspond to the three emission peaks of the LED light source equipped with the Raman spectrometer, and the peak at 605 nm is labeled as peak O (Orange). For the reflection spectra from the MoS$_2$ nanofilms, three additional peaks appear. The two lowest energy peaks (~1.79 and 1.92 eV) correspond to the A and B excitons associated with interband transitions between the maximum of the split valence bands induced by spin-orbit coupling and the minimum of the conduction band at the K and K′ point in the Brillouin zone.[57] The broad response at higher photon energy (~2.36 eV) consists of six nearly degenerate exciton states made from transition occurs between the K and Γ points of the BZ and is denoted by C in Figure 1e.[57] The above assignment is also consistent with the results of the differential reflectance spectra (Figure S3). The positions and intensities of peaks A, B, and C all change as the layer number of MoS$_2$ increases. Specifically, the positions of all three peaks are red-shifted as the MoS$_2$ thickness increases due to quantum-confinement[58] and optical inteference.[24] However, the intensities of the three peaks have different evolving trends. For 1-3L MoS$_2$, peaks A and B are barely visible due to interlayer charge transfer in the utra-thin MoS$_2$-Au hybrid structure as we previously reported.[51] Beyond 3L, the intensities of both peaks decrease monotonously with the increase of sample thickness. In contrast, the intensity of peak C increases monotonously as shown in Figure 1f. More importantly, when the thickness is greater than 5L, the intensity of peak C increases almost linearly with a slope of 0.052, suggesting that it can be used for accurate layer number determination, especially for thicker MoS$_2$ nanofilms. The thickness dependence of the reflected light intensity will be discussed in detail later.

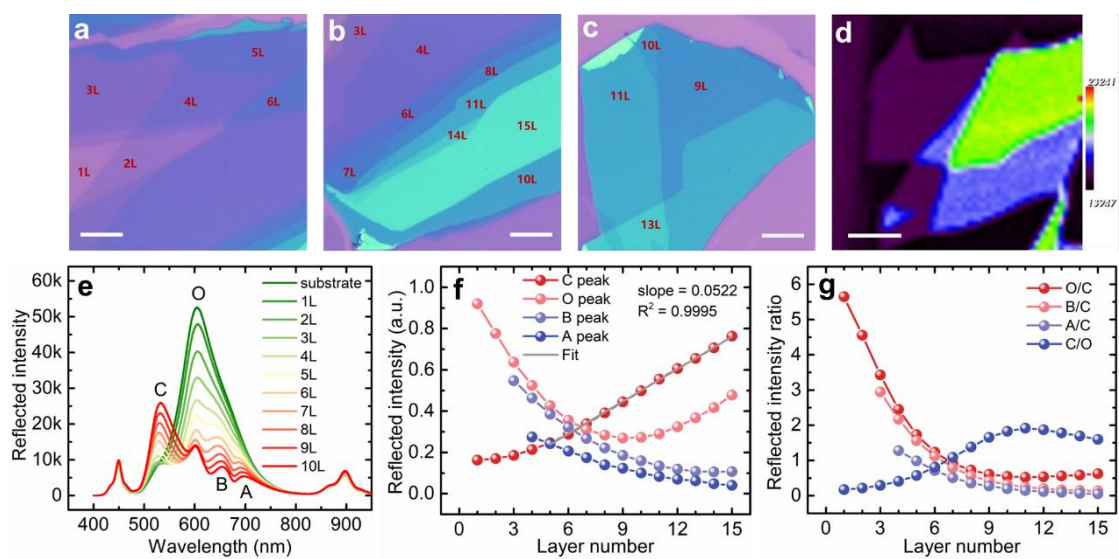

**Figure 1**. a-c) Optical microscope images of Au-assisted exfoliated large area 1-15L MoS$_2$ nanofilms. The scale bar is 10 μm. d) White light reflection spectra mapping at 605 nm of a MoS$_2$ nanoflake with regions of different thicknesses. e) Layer-dependent white light reflection spectra of 1-10L MoS$_2$. f) Reflection peak intensities of A, B, C, and O peaks collected from 1-15L MoS$_2$ normalized by the O peak intensity of the substrate reflection spectra. g) Reflection peak intensity ratios of O/C, B/C, A/C, and C/O.

For 1-5L MoS$_2$, although the intensity of reflection peak C increases monotonously, the increase is still too slow to be conducive for accurate thickness determination. We noticed that when the thickness of MoS$_2$ increases, the positions of the O emission peak of the LED light source do not change, while the intensity decreases significantly first (1-9L) and then increases (> 9L) with the increase of the MoS$_2$ thickness. And it can be used to accurately determine the layer number of 1-7L MoS$_2$, which cannot be easily distinguished by peak C. Once the 1-7L MoS$_2$ is determined by peak O, the slope of the linear increase region in the intensity of peak C can then be determined and used to accurately identify the layer number of thicker MoS$_2$ films. Moreover, we note that the peak intensity ratio between peak O and peak C (O/C) shown in Figure 1g can also be used to accurately determine the thickness of 1-9L MoS$_2$, whose distinguishing effect far exceeds the Raman frequency difference method shown later (Figure 3b).

To test the thickness limit that can be determined using this method, we probe its sensitivity by measuring thicker MoS$_2$ nanoflakes (16-35L, confirmed by AFM measurements in the Supporting Information) as shown in **Figure 2**a-c. Based on the analysis of the thickness determination of 1-15L MoS$_2$, we now only focus on the intensities of peaks C and O in the following discussion. As shown in Figure 2e, for the 16-35L MoS$_2$, the linear relationship between the intensity of peak C and the sample thickness remains. Therefore, for 5-35L MoS$_2$, its thickness can be accurately determined by the intensity of the peak C. As for the peak O, there is a small jump in its intensity at 22L. This is due to the fact that the reflection spectra of the 22L MoS$_2$ has a flat band in the 540-575 nm range and the position of peak O is obviously shifted before and after this thickness (see Figure S4 in the Supporting Information for details).

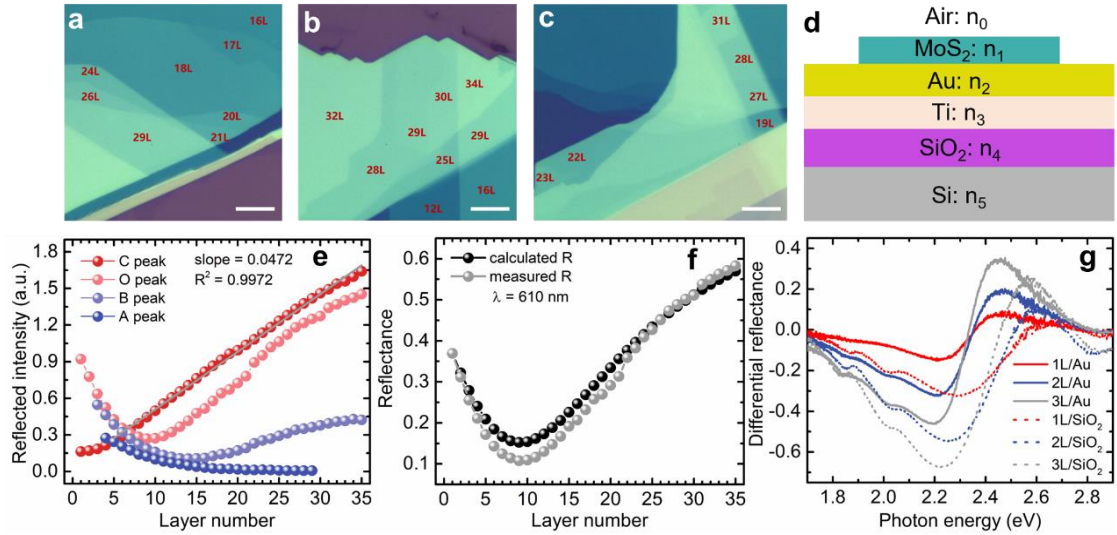

**Figure 2**. a-c) Optical microscope images of Au-assisted exfoliated large area 16-35L MoS$_2$ nanoflakes. The scale bar is 10 µm. (d) Schematic diagram of the quintuple layer structure consisting of MoS$_2$, Au, Ti, SiO$_2$, and Si with normal white light incidence from air. e) Normalized reflection peak intensities of A, B, C, and O peaks collected from 1-35L MoS$_2$. f) Calculated and measured layer dependent reflectance at 610 nm of the MoS$_2$/Au/Ti/SiO$_2$/Si quintuple layer structure in d). g) Layer dependent differential reflectance spectra of 1-3L MoS$_2$ supported on the Au/Ti/SiO$_2$/Si substrate and the SiO$_2$/Si substrate.

To explain the thickness dependence of the reflected light intensity, the reflectance of the MoS$_2$/Au/Ti/SiO$_2$/Si quintuple layer system (shown in Figure 2d) has been calculated based on Fresnel's theory. On top of the unsupported thin film, which can be equivalent to a double-layer system with membrane on semi-infinite air layer,[59] the reflectance of the quintuple layer system in our work can be calculated using the recursive method (see **Note S1** in the Supporting Information for details of calculation)

$$R(\lambda) = |r(\lambda)|^2 \tag{1}$$

$$r(\lambda) = r_a/r_b \tag{2}$$

$$r_a = r_1 e^{2i\beta_3} + r_2 e^{2i(\beta_3-\beta_1)} + r_3 e^{2i(\beta_3-\beta_2-\beta_1)} + r_4 e^{-2i(\beta_1+\beta_2)} + r_5 e^{-2i(\beta_1+\beta_2+\beta_4)} +$$

$$r_1 r_2 r_3 e^{2i(\beta_3-\beta_2)} + r_1 r_2 r_4 e^{-2i\beta_2} + r_1 r_2 r_5 e^{-2i(\beta_2+\beta_4)} + r_1 r_3 r_4 + r_1 r_3 r_5 e^{-2i\beta_4} +$$

$$r_1 r_4 r_5 e^{-2i(\beta_4-\beta_3)} + r_1 r_2 r_3 r_4 r_5 e^{-2i(\beta_4-\beta_3+\beta_2)} + r_2 r_3 r_4 e^{-2i\beta_1} + r_2 r_3 r_5 e^{-2i(\beta_1+\beta_4)} +$$

$$r_2 r_4 r_5 e^{-2i(\beta_1-\beta_3+\beta_4)} + r_3 r_4 r_5 e^{-2i(\beta_1+\beta_2-\beta_3+\beta_4)} \tag{3}$$

$$r_b = e^{2i\beta_3} + r_1 r_2 e^{2i(\beta_3-\beta_1)} + r_1 r_3 e^{2i(\beta_3-\beta_2-\beta_1)} + r_1 r_4 e^{-2i(\beta_1+\beta_2)} + r_1 r_5 e^{-2i(\beta_1+\beta_2+\beta_4)} +$$

$$r_2 r_3 e^{2i(\beta_3-\beta_2)} + r_2 r_4 e^{-2i\beta_2} + r_2 r_5 e^{-2i(\beta_2+\beta_4)} + r_3 r_4 + r_3 r_5 e^{-2i\beta_4} + r_4 r_5 e^{-2i(\beta_4-\beta_3)} +$$

$$r_2 r_3 r_4 r_5 e^{-2i(\beta_4-\beta_3+\beta_2)} + r_1 r_2 r_3 r_4 e^{-2i\beta_1} + r_1 r_2 r_3 r_5 e^{-2i(\beta_1+\beta_4)} + r_1 r_2 r_4 r_5 e^{-2i(\beta_1-\beta_3+\beta_4)} +$$

$$r_1 r_3 r_4 r_5 e^{-2i(\beta_1+\beta_2-\beta_3+\beta_4)} \tag{4}$$

where $r_i=(n_{i-1}-n_i)/(n_{i-1}+n_i)$ is the Fresnel reflection coefficients at the interface from medium $i$-1 to $i$; the indices are assigned as air (0), $MoS_2$ (1), Au (2), Ti (3), $SiO_2$ (4), and Si (5). $n_i$ is the complex refractive index of the $i$th layer with $n_0 = 1$. $\beta_i = 2\pi n_i d_i/\lambda$ is the phase factor, representing the phase differences through the whole medium $i$, where $d_i$ is the thickness of medium $i$. The thickness of the $MoS_2$ nanofilm is estimated by $d = N\Delta d$, where $N$ represents the layer number and $\Delta d$ is the thickness of monolayer $MoS_2$ (0.615 nm).[60] The material thicknesses and wavelength dependent refractive indices of Au ($d_2$, $n_2$), Ti ($d_3$, $n_3$), $SiO_2$ ($d_4$, $n_4$), Si (semi-infinite, $n_5$), $MoS_2$ and $WS_2$ (bulk, $n_1$) were determined by spectroscopic ellipsometry measurement as we previously reported[51] and were also shown in Figure S5 and Table S1. Given that the positions of excitonic peaks (A, B, and C) change significantly with the $MoS_2$ thickness, while the position of the peak O is almost unchanged, we calculated layer-dependent reflectance of our system at 610 nm as shown in Figure 2f, which is in good agreement with the measured reflectance. The slight difference between the two may be due to the layer-dependent dielectric function of few layer $MoS_2$ revealed by our layer-dependent differential reflectance spectra in Figure 1g and Figure S6, which may be slightly different from the bulk value used in our calculations. Nevertheless, the overall consistency indicates that the layer dependence of reflected peak intensity mainly originates from classical interference effects.

We want to note that many systematic studies on the layer-dependent physical properties can be carried out once the layer number of the Au-assisted exfoliated 2D films is determined. For example (and most conveniently), the layer-dependent dielectric function can be obtained from the measured differential reflectance spectra by the Kramers-Kronig constrained analysis.[61] The differential reflectance spectra of the 1-3L $MoS_2$ supported on both the Au substrate and the $SiO_2$/Si substrate are shown in Figure 2g to get a preliminary understanding of the characteristics of 2D films-Au hybrid system's layer-dependent dielectric function. Since the Au substrate can provide large strain and promote the interlayer charge transfer at the Au/2D films interface,[51-52] the optical properties of the 2D films-Au hybrid system are different from the intrinsic ones as revealed in Figure 2g and from our previous ellipsometric study on the 1L $MoS_2$-Au hybrid system.[51] Nevertheless, the systematic study of the layer-dependent dielectric function of the hybrid system still requires future efforts, which is not the focus of this article. In addition, the Au substrate facilitates the measurement of electron-based characterizations, such as scanning tunneling microscopy (STM),[15] angle-resolved photoelectron spectroscopy (ARPES),[62] and conductive atomic force microscope (CAFM),[63] which will in turn promote the study of layer-dependent electronic band structures and electrical transport properties.

To further verify the reliability of our method to determine the layer number, we benchmark it against the commonly used Raman spectroscopy. PL spectroscopy is not used for benchmarking as the PL signal is quenched for $MoS_2$ on Au due to interlayer charge transfer (see the thickness-dependent PL spectra in Figure S7). **Figure 3**a shows the layer-dependent Raman spectra of 1-15L $MoS_2$. In principle, the frequency difference between the $E_{2g}^1$ and $A_{1g}$ modes is anticipated to increase monotonously with the layer number and can be used to distinguish 1-6L $MoS_2$.[27] In the case of $MoS_2$ on Au, however, the abnormally large frequency difference of 1L $MoS_2$ is due to the large red shift of the $E_{2g}^1$ mode caused by the large strain provided by the Au film as we have previously reported.[51] Nevertheless, the frequency differences obtained from $MoS_2$ on Au are consistent with the values from $MoS_2$ on $SiO_2$/Si for 2-6L $MoS_2$. This gives the same layer numbers as our reflection peak intensity method. Figure 3c shows the layer-dependent Raman intensities of $MoS_2$ and Si (~ 520 cm$^{-1}$). The Raman intensities of the $A_{1g}$ and $E_{2g}^1$ modes both reach the maximum value at 9 L and then drop with increasing thickness due to optical interference; the layer number where the maximum Raman intensity appears depends on the specific optical structure of the multilayer system.[64] The Raman intensity ratios of the $MoS_2$ $A_{1g}$ and $E^1_{2g}$ modes to the Si mode are also plotted in Figure 3d. Compared with the frequency difference in Figure 3b and the intensities in Figure 3c, the Raman intensity ratio presents a clearer linear dependence on the layer number in a wider thickness range (~15L). Particularly, the distinguishing effect of $A_{1g}/Si$ is better than that of $E^1_{2g}/Si$ due to its better linearity and larger slope as revealed by the linear fits in Figure 3d. However, as the layer number of MoS$_2$ further increases, the fluctuation and saturation of data points make the Raman based metrics no longer effective as shown in the yellow regions in Figure 3d. In fact, the Raman frequency difference and intensities are usually effective only in distinguishing 1-6L MoS$_2$ with high accuracy. In comparison, our reflected peak intensity method exhibits great advantages in a much larger layer number identification range (at least 30L) with higher accuracy for thicker films.

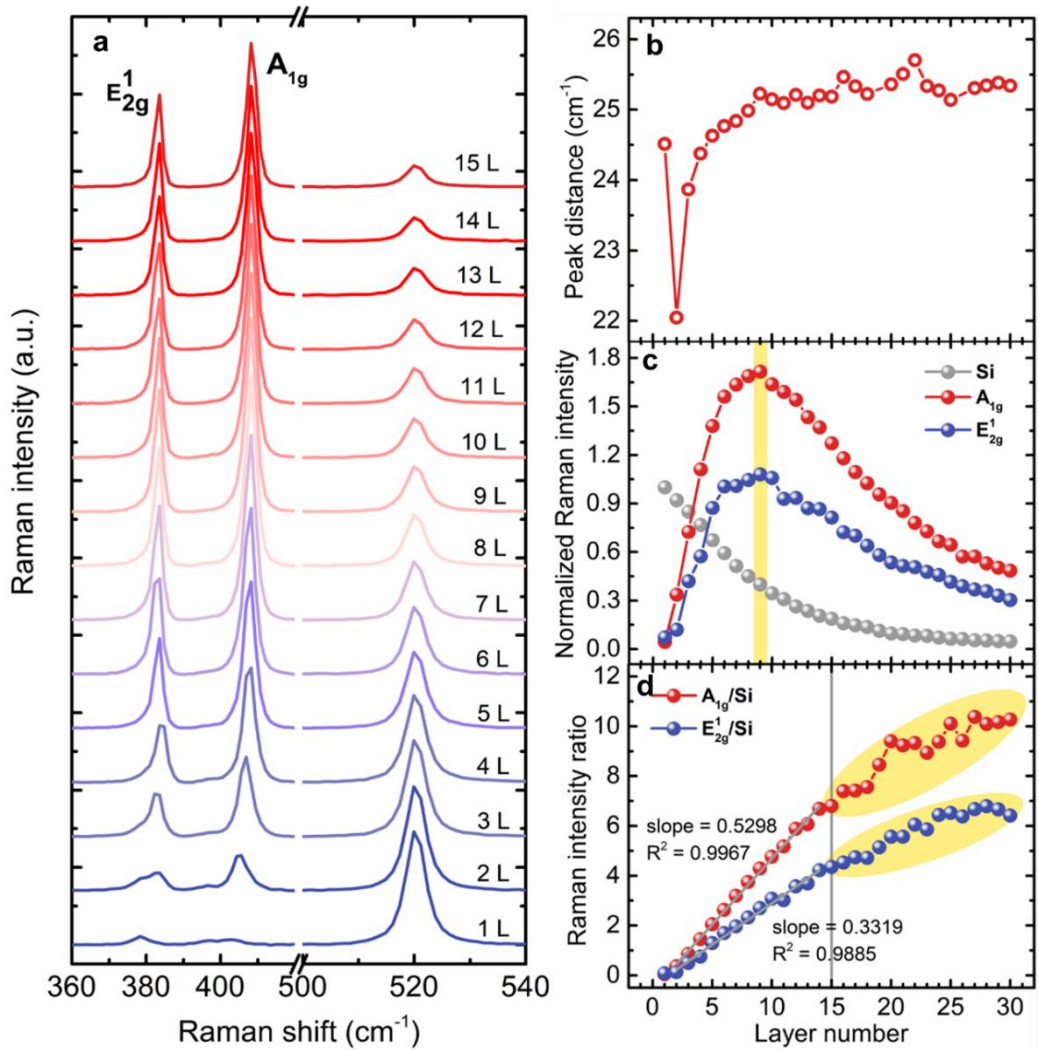

**Figure 3**. a) Layer-dependent Raman spectra of 1-15L MoS$_2$. b) Layer-dependent

frequency difference of 1-30L MoS$_2$. c) Layer-dependent Raman intensities of the $A_{1g}$ and $E_{2g}^1$ modes of 1-30L MoS$_2$ as well as the Si mode normalized by the Raman intensity of the Si mode of the substrate. d) Layer-dependent Raman intensity ratios of the $A_{1g}/Si$ and $E_{2g}^1/Si$.

The effectiveness of this method should in principle apply to other TMDCs, so we extended this method to WS$_2$. **Figure 4**a-c shows the OM images of Au-assisted exfoliated large area 1-17L WS$_2$ nanofilms (confirmed by AFM measurements in the Supporting Information). The layer-dependent WLRS of 1-10L WS$_2$ is shown in Figure 4e (see Figure S9 for reflection spectra of 1-18L WS$_2$). Different from those of MoS$_2$, there are only two distinct sample peaks in the reflection spectra of WS$_2$ as one of the sample peaks (peak B) overlaps with O peak .[65-66] Other than that, the peak intensity and peak position both change significantly with the thickness of WS$_2$ similar to MoS$_2$. The normalized reflection peak intensities of A, C, and O peaks collected from 1-18L WS$_2$ are shown in Figure 4f. The intensity of peak A decreases monotonously while that of peak C increases monotonously with the increase of sample thickness and these two peaks are almost indistinguishable for monolayer and bilayer WS$_2$. Notably, when the thickness of WS$_2$ nanofilm is greater than 7L, the intensity of peak C increases almost linearly with a slightly larger slope of 0.0527 than MoS$_2$ (0.0522). As for peak O, the intensity decreases first and then increases with a minimum value at 14L, which is larger that of MoS$_2$ (9L) due to the different optical constants of the two materials (Figure S5). Moreover, the intensity of peak O changes significantly for 1-9L WS$_2$. Therefore, for WS$_2$ with a thickness of 1-7L and more than 7L, the number of layers can be accurately determined by the intensity of peak O and peak C, respectively. We also calculated the layer-dependent reflectance of WS$_2$ on the Au substrate at 605 nm, and once again it is in good agreement with the experiment as shown in Figure 4g, which further reveals the reliability of our WLRS based method for thickness determination.

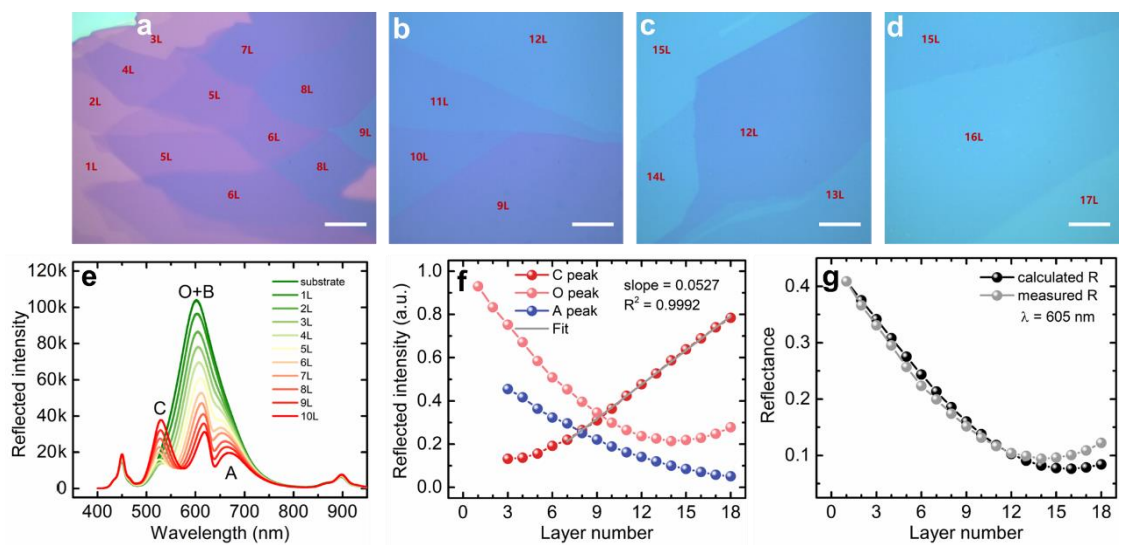

**Figure 4**. a-d) Optical microscope images of Au-assisted exfoliated large area 1-17L WS$_2$ nanoflakes. The scale bar is 10 µm. e) Layer-dependent white light reflection spectra of 1-10L WS$_2$. f) Normalized reflection peak intensities of A, C, and O peaks collected from 1-18L WS$_2$. g) Calculated and measured layer dependent reflectance at 605 nm of the WS$_2$/Au/Ti/SiO$_2$/Si five-layer structure.

Although the emerging Au-assisted exfoliation technique has shown great advantages in the preparation of large-area single-crystalline 2D thin films including monolayers, tape-based mechanical exfoliation remains an important method owing to its feasibility and flexibility for dry transfer after exfoliation. Therefore, we also made a theoretical prediction on the thickness determination on tape-exfoliated and tranferred TMDCs films. The calculated layer-dependent reflectances of MoS$_2$ and WS$_2$ supported on the SiO$_2$ (272.7 and 310.4 nm)/Si and Au/Ti/SiO$_2$/Si substrates at 605 nm are shown in **Figure 5**. Obviously, the WLRS based method is also applicable to traditional tape-exfoliated films as the principle of the method is the classical interference effect, and this method retains broad range of application in different situations. However, compared with the SiO$_2$/Si substrate, the Au substrate can determine the thickness of a thicker film and it is easy to distinguish ultra-thin few layer films, because peak G has a large drop interval at the beginning. In addition, the difference in thickness dependent reflectivity of MoS$_2$ and WS$_2$ is obvious for the Au substrate (not caused by slight differences in oxide layer thickness, Figure S10), while the two materials are almost the same for the SiO$_2$/Si substrate. As for the two SiO$_2$/Si substrates we chose, the one with 270 nm oxide layer is more suitable for thickness judgment. Because for the substrate with 310 nm oxide layer, peak G has a very narrow drop range, which is not conducive for the distinction of few layer samples. In short, regarding the effectiveness of the WLRS method to determine the layer number, the Au substrate is better than the SiO$_2$ (270 nm)/Si substrate and the SiO$_2$ (270 nm)/Si substrate is better than the SiO$_2$ (310 nm)/Si substrate.

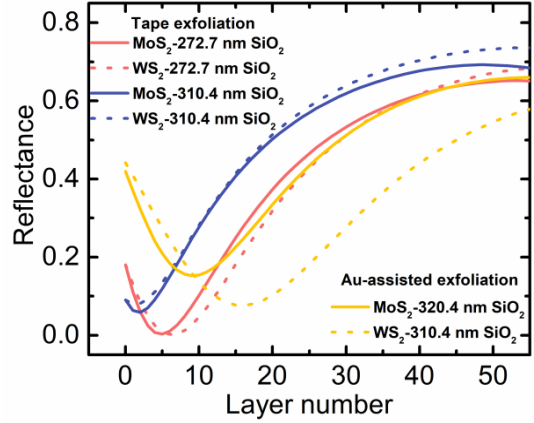

**Figure 2**. Calculated layer dependent reflectance at 605 nm of MoS$_2$ and WS$_2$ supported on the Au/Ti/SiO$_2$/Si substrate and SiO$_2$/Si substrates with different oxide layer thicknesses.

## 3. Conclusions

In summary, a simple, fast yet reliable optical method is proposed to determine the layer number of MoS$_2$ and WS$_2$ films prepared by Au-assisted mechanical exfoliation, which is an emerging technique to obtain large area and high quality TMDCs. The method takes advantage of the reflection intensities of the sample peaks (A, B, and C) and the 605 nm peak of the white LED (peak O) which exhibit clear evolving patterns with the layer number. Specifically, the intensity of reflection peaks O and C can be used as a clear indicator for precise identification of layer number of ultra-thin and thicker films, respectively. We have benchmarked this white light reflection spectra method with other commonly used techniques such as AFM and Raman spectroscopy, finding an excellent agreement. Calculations based on Fresnel theory show that this method is based on optical interference; it thus has a large identification range of more than 30 layers and is also applicable for traditional tape-exfoliated TMDCs films. The simple and robust optical identification method will facilitate the fundamental study on layer-dependent physical properties and demonstration of proof-of-concepts devices based on TMDCs. Moreover, the spectroscopy-based method is readily integrated with Raman and PL measurements for the simultaneous study of other physical properties.

## 4. Experimental details and methods

*Gold-Assisted Exfoliation:* The SiO$_2$/Si wafers were sonicated for 25 min each in acetone, isopropanol, ethanol and then dried with a nitrogen gun. PVD metal films were prepared with a DC magnetron sputtering system (Kurt J. Lesker CMS-A) at room temperature under a 5 SCCM Ar flow at partial pressure of $10^{-3}$ Torr. A Ti adhesion layer was first deposited on SiO$_2$/Si wafer, and then Au film was deposited. Both metal layers were deposited for 1 min at a power of 100 W (for Ti) and 60 W (for Au), respectively. MoS$_2$ and WS$_2$ were cleaved from bulk crystals (purchased from Shanghai ONWAY technology Co., Ltd) with scotch-tape and then pressed onto the freshly prepared Au substrate. The stack was annealed on a hotplate in ambient conditions at 180 °C for 60 s and taken from the hotplate to cool down for 10–20 s before peeling the tape. Large-area monolayer and few layer flakes can be obtained after the tape was removed.

*Raman, Photoluminescence and Reflection Spectra Measurements:* Raman, photoluminescence spectra and white light reflection spectra were collected on a Renishaw InVia Raman Microscope system. When collecting the white light reflection spectra, a white LED source was used and the edge filter in the signal light path was removed. The light intensity of the white light source can be automatically and accurately controlled, so the experimental results of different batches have excellent repeatability. Raman and photoluminescence spectra were obtained in a back scattering geometry with a 532 nm excitation laser. A 100× objective lens with NA = 0.85 and a 1800 lines/mm grating were used to collect Raman, photoluminescence signals and reflected white light.

*Characterization of Samples and Substrates:* The thicknesses of the MoS$_2$ and WS$_2$ films were determined by AFM (Bruker Dimension ICON). The dielectric functions of bulk MoS$_2$ and WS$_2$ and the actual thicknesses of SiO$_2$, Ti, and Au layers as well as their dielectric functions were determined by spectroscopic ellipsometry measurements (J. A. Woollam Co., Inc.)

**Supporting Information**:
The Supporting Information to this article is available online or from the author.

**Acknowledgements:**

This work was supported by the Natural Science Foundation of Guangdong Province (grant no. 2020A1515010885) and the Shenzhen Science and Technology Innovation Programs (grant no JCYJ20190806142614541).

**Conflict of Interest:**

The authors declare no conflict of interest.

# Supporting Information:

# Reliable and Broad-range Thickness Determination of for Au-assisted Exfoliated Large Area $MoS_2$ and $WS_2$ Using Reflection Spectroscopy

Bo Zou[1], Yu Zhou[1], Yan Zhou[2], Yanyan Wu[1], Yang He[1], Xiaonan Wang[1], Jinfeng Yang[1], Lianghui Zhang[1], Yuxiang Chen[1], Shi Zhou[3], Huaixin Guo[4], Huarui Sun[1],[*]

[1] *School of Science and Ministry of Industry and Information Technology Key Laboratory of Micro-Nano Optoelectronic Information System*

*Harbin Institute of Technology Shenzhen, 518055, P. R. China*

*Collaborative Innovation Center of Extreme Optics, Shanxi University, Taiyuan, Shanxi 030006, People's Republic of China*

[2] *State Key Laboratory of Superlattices and Microstructures, Institute of Semiconductors, Chinese Academy of Sciences*

*Beijing 100083, P. R. China*

[3] University of Science and Technology of China, Hefei 230026, P. R. China

[4] Science and Technology on Monolithic Integrated Circuits and Modules Laboratory, Nanjing Electronic Devices Institute, Nanjing 210016, P.R. China

[*]Corresponding author. Email address: huarui.sun@hit.edu.cn

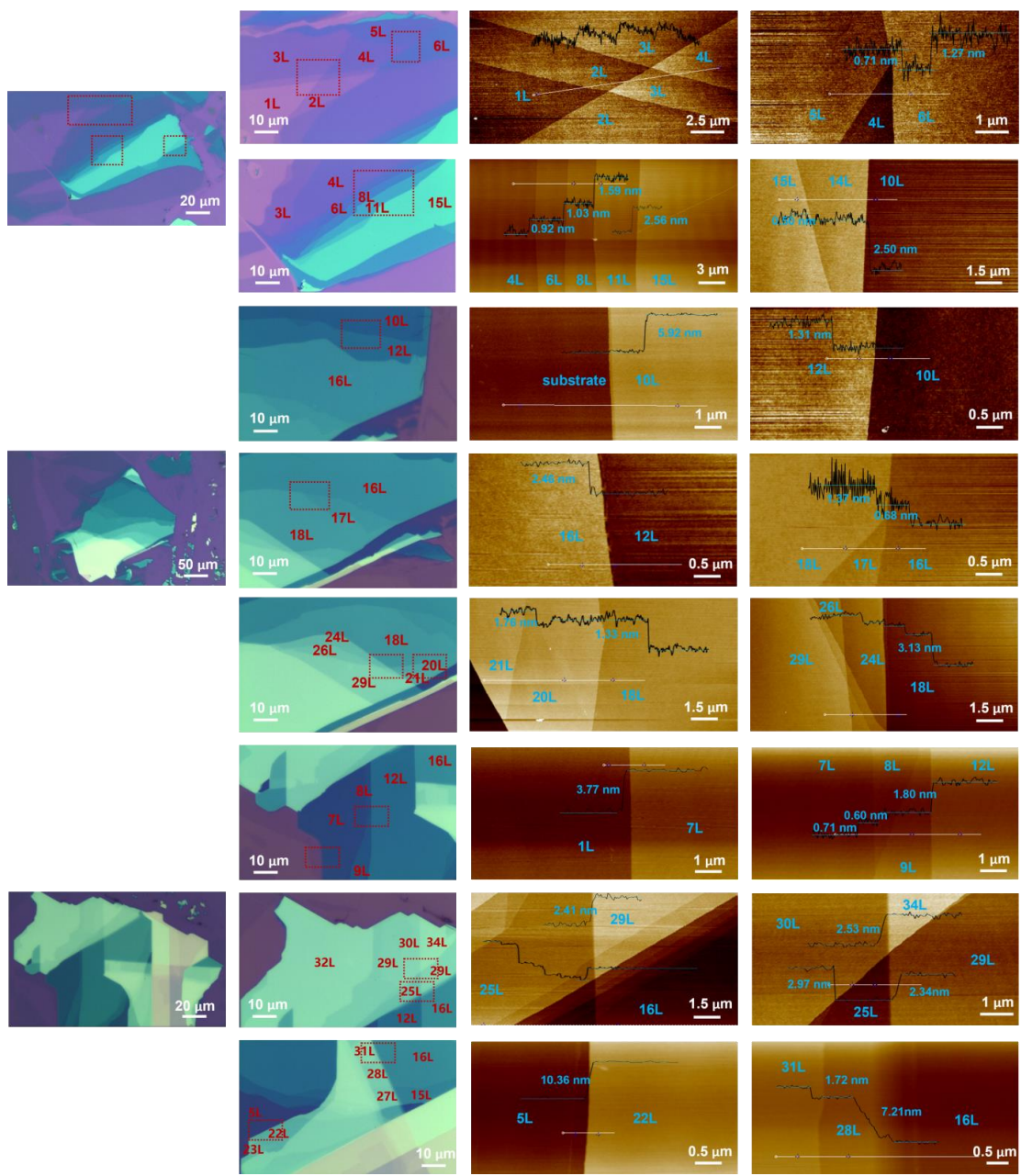

**Figure S1**. AFM thickness measurements of three $MoS_2$ flakes with regions of 1-35L.

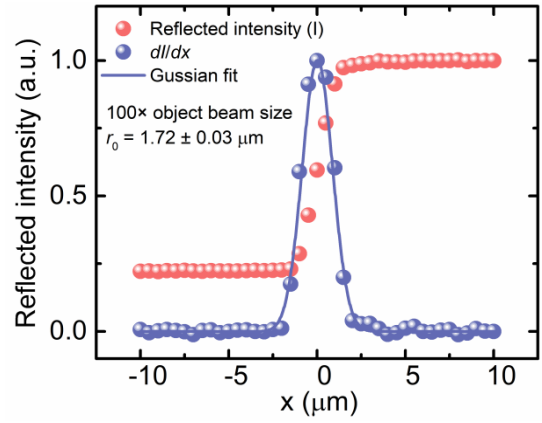

Figure S2. Knife-edge measurement of white light spot radius focused through a 100× objective. The reflected white light intensity (red dots) and Gaussian fitted profile (purple solid line) of the beam as a function of the beam position, giving a spot radius $r_0 = 1.72 \pm 0.03$ μm.

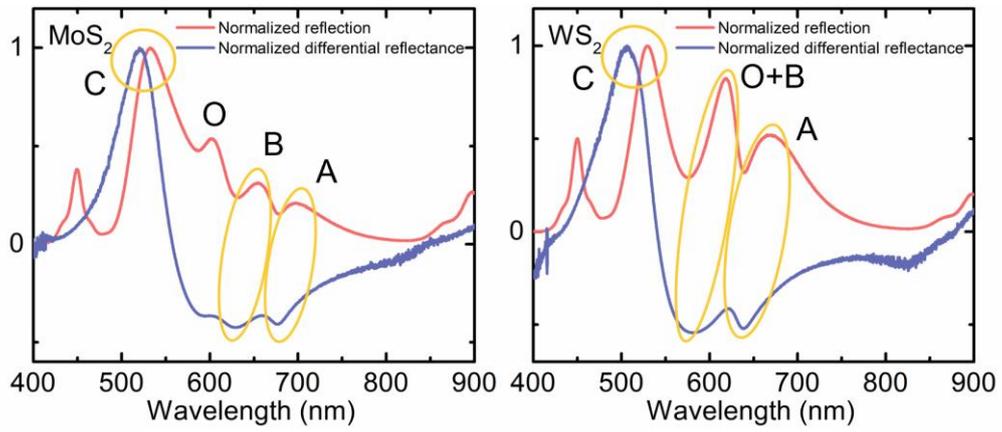

Figure S3. Normalized reflection spectra and differential reflectance spectra of (a) $MoS_2$ and (b) $WS_2$. The comparison of the two types of spectra clarifies the origin of the reflection peaks.

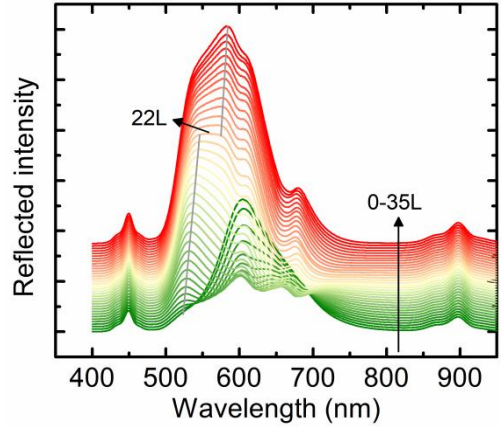

Figure S4. Layer-dependent reflection spectra of 1-35L MoS$_2$. The reflection spectrum of 22L has a flat interval from 540-575 nm and the small jump of the G peak intensity on the 22L in Figure 2e is due to the shift of the G peak position before and after the 22L as shown by the gray lines.

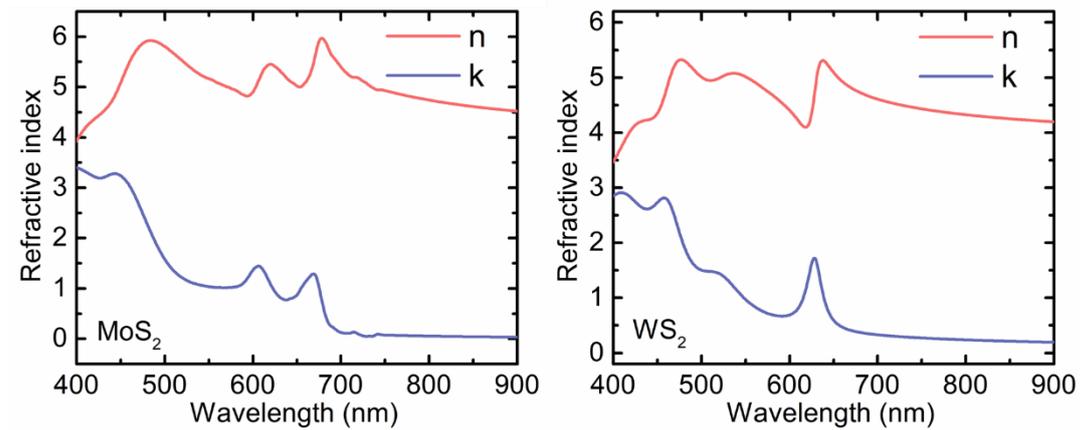

**Figure S5**. Refractive index of bulk (a) MoS$_2$ and (b) WS$_2$ in the range of 210-1000 nm measured by spectroscopic ellipsometry.

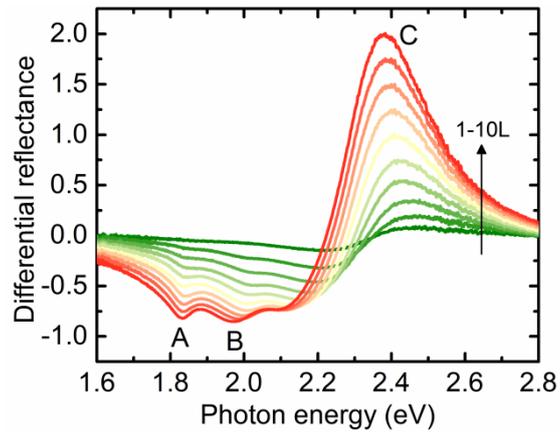

**Figure S6**. Layer-dependent differential reflectance spectra of 1-10L MoS$_2$.

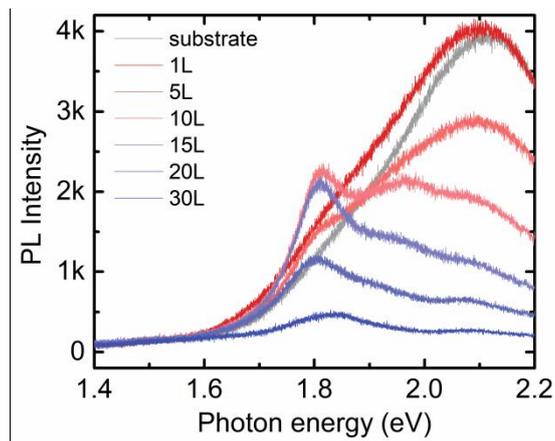

**Figure S7**. Thickness-dependent PL spectra of 1-10L MoS$_2$.

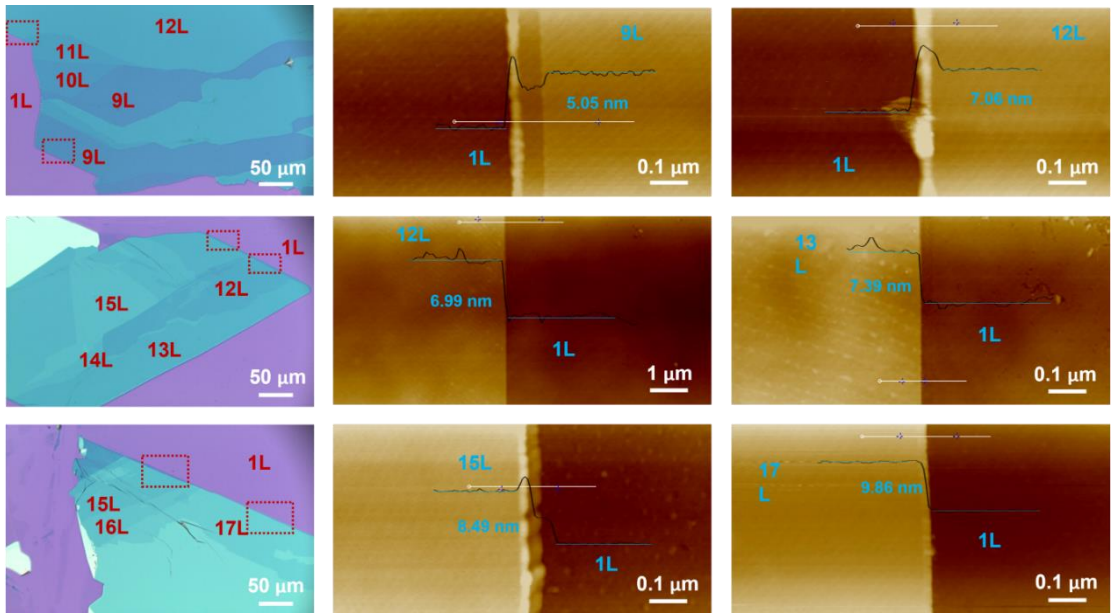

Figure S8. AFM thickness measurements of WS$_2$ flakes.

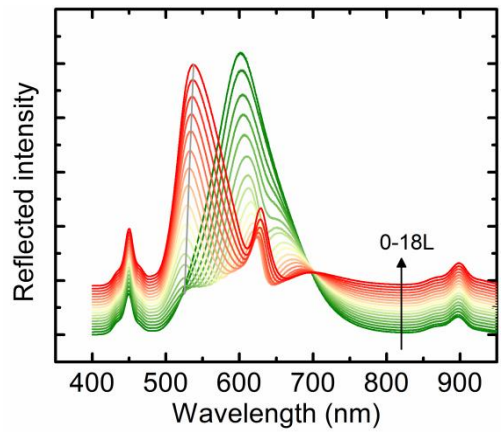

Figure S9. Layer-dependent reflection spectra of 1-18L WS$_2$.

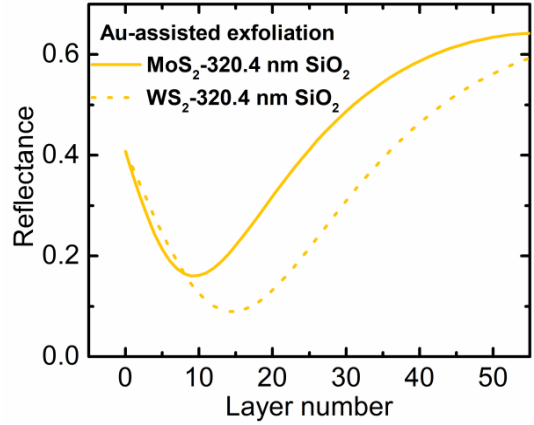

**Figure S10.** Calculated layer dependent reflectance of $MoS_2$ and $WS_2$ supported on the same substrate. Obviously, compared with the $SiO_2$/Si substrate substrate, for the Au substrate used in our experiment, the difference in reflectance between the two materials is magnified.

**Table S1. Wavelength-dependent complex refractive indices used in the calculation of reflectance.**

|  | 600 nm | 605 nm | 610 nm |
|---|---|---|---|
| **$MoS_2$ (bulk)** | 4.900-1.368i | - | 5.268-1.403i |
| **$WS_2$ (bulk)** | 4.405-0.690i | 4.314-0.733i | - |
| **Au (10.46 nm)** | - | 0.411-3.948i | 0.414-3.996i |
| **Ti (6.52 nm)** | - | 2.203-2.643i | 2.218-2.657i |
| **$SiO_2$ (272.7 nm)** | 1.458 | - | - |
| **$SiO_2$ (310.4 nm)** | 1.457 | 1.457 | - |
| **$SiO_2$ (320.4 nm)** | - | - | 1.463 |
| **Si** | 3.940-0.018i | 3.929-0.018i | 3.918-0.017i |

## Supporting Note S1

### Calculation of the reflectance of the quintuple layer system

The reflectance of the quintuple layer system can be calculated using the recursive method based on the reflectance of the bilayer system.

For an unsupported thin film (can be regarded as a bilayer system with thin film and semi-infinite air layer), the normal incidence reflectance $R''(\lambda)$ is given by[1]

$$R''(\lambda) = |r''(\lambda)|^2 \tag{S1}$$

the reflection coefficient is

$$r''(\lambda) = \frac{r_1 + r_2 e^{-2i\beta_1}}{1 + r_1 r_2 e^{-2i\beta_1}} \tag{S2}$$

For trilayer system, such as thin film/SiO$_2$/Si system, the reflection coefficient $r'''(\lambda)$ can be iteratively obtained by the method shown below on the basis of $r''(\lambda)$

$$r'''(\lambda) = \frac{r_1 + \left(\frac{r_2 + r_3 e^{-2i\beta_2}}{1 + r_2 r_3 e^{-2i\beta_2}}\right) e^{-2i\beta_1}}{1 + r_1 \left(\frac{r_2 + r_3 e^{-2i\beta_2}}{1 + r_2 r_3 e^{-2i\beta_2}}\right) e^{-2i\beta_1}} = \frac{r_1 e^{2i\beta_1} + r_2 + r_3 e^{-2i\beta_2} + r_1 r_2 r_3 e^{2i(\beta_1 - \beta_2)}}{e^{2i\beta_1} + r_1 r_2 + r_1 r_3 e^{-2i\beta_2} + r_2 r_3 e^{2i(\beta_1 - \beta_2)}} \tag{S3}$$

Modification of the equation S3 (multiplying the numerator and denominator by the factor $e^{-i(\beta_1 - \beta_2)}$ at the same time), the common form in the literature[2] can be obtained.

Similarly, the reflection coefficient of the quadralayer system $r''''(\lambda)$ is

$$r''''(\lambda) = \frac{r_1 + \left(\frac{r_2 e^{2i\beta_2} + r_3 + r_4 e^{-2i\beta_3} + r_2 r_3 r_4 e^{2i(\beta_2 - \beta_3)}}{e^{2i\beta_2} + r_2 r_3 + r_2 r_4 e^{-2i\beta_3} + r_3 r_4 e^{2i(\beta_2 - \beta_3)}}\right) e^{-2i\beta_1}}{1 + r_1 \left(\frac{r_2 e^{2i\beta_2} + r_3 + r_4 e^{-2i\beta_3} + r_2 r_3 r_4 e^{2i(\beta_2 - \beta_3)}}{e^{2i\beta_2} + r_2 r_3 + r_2 r_4 e^{-2i\beta_3} + r_3 r_4 e^{2i(\beta_2 - \beta_3)}}\right) e^{-2i\beta_1}} =$$

$$\frac{r_1 e^{2i\beta_2} + r_2 e^{2i(\beta_2 - \beta_1)} + r_3 e^{-2i\beta_1} + r_4 e^{-2i(\beta_1 + \beta_3)} + r_1 r_2 r_3 + r_1 r_2 r_4 e^{-2i\beta_3} + r_1 r_3 r_4 e^{2i(\beta_2 - \beta_3)} + r_2 r_3 r_4 e^{-2i(\beta_1 - \beta_2 + \beta_3)}}{e^{2i\beta_2} + r_1 r_2 e^{2i(\beta_2 - \beta_1)} + r_1 r_3 e^{-2i\beta_1} + r_1 r_4 e^{-2i(\beta_1 + \beta_3)} + r_2 r_3 + r_2 r_4 e^{-2i\beta_3} + r_3 r_4 e^{2i(\beta_2 - \beta_3)} + r_1 r_2 r_3 r_4 e^{-2i(\beta_1 - \beta_2 + \beta_3)}}$$

$$\tag{S4}$$

For quintlayer system, the reflection coefficient $r'''''(\lambda)$ is

$$r'''''(\lambda) =$$

$$\frac{r_1 + \left(\frac{r_2 e^{2i\beta_3} + r_3 e^{2i(\beta_3 - \beta_2)} + r_4 e^{-2i\beta_2} + r_5 e^{-2i(\beta_2 + \beta_4)} + r_2 r_3 r_4 + r_2 r_3 r_5 e^{-2i\beta_4} + r_2 r_4 r_5 e^{2i(\beta_3 - \beta_4)} + r_3 r_4 r_5 e^{-2i(\beta_2 - \beta_3 + \beta_4)}}{e^{2i\beta_3} + r_2 r_3 e^{2i(\beta_3 - \beta_2)} + r_2 r_4 e^{-2i\beta_2} + r_2 r_5 e^{-2i(\beta_2 + \beta_4)} + r_3 r_4 + r_3 r_5 e^{-2i\beta_4} + r_4 r_5 e^{2i(\beta_3 - \beta_4)} + r_2 r_3 r_4 r_5 e^{-2i(\beta_2 - \beta_3 + \beta_4)}}\right) e^{-2i\beta_1}}{1 + r_1 \left(\frac{r_2 e^{2i\beta_3} + r_3 e^{2i(\beta_3 - \beta_2)} + r_4 e^{-2i\beta_2} + r_5 e^{-2i(\beta_2 + \beta_4)} + r_2 r_3 r_4 + r_2 r_3 r_5 e^{-2i\beta_4} + r_2 r_4 r_5 e^{2i(\beta_3 - \beta_4)} + r_3 r_4 r_5 e^{-2i(\beta_2 - \beta_3 + \beta_4)}}{e^{2i\beta_3} + r_2 r_3 e^{2i(\beta_3 - \beta_2)} + r_2 r_4 e^{-2i\beta_2} + r_2 r_5 e^{-2i(\beta_2 + \beta_4)} + r_3 r_4 + r_3 r_5 e^{-2i\beta_4} + r_4 r_5 e^{2i(\beta_3 - \beta_4)} + r_2 r_3 r_4 r_5 e^{-2i(\beta_2 - \beta_3 + \beta_4)}}\right) e^{-2i\beta_1}} =$$

$$\frac{r_a}{r_b} \tag{S5}$$

42  $r_a = r_1 e^{2i\beta_3} + r_2 e^{2i(\beta_3-\beta_1)} + r_3 e^{2i(\beta_3-\beta_2-\beta_1)} + r_4 e^{-2i(\beta_1+\beta_2)} + r_5 e^{-2i(\beta_1+\beta_2+\beta_4)} +$

43  $r_1 r_2 r_3 e^{2i(\beta_3-\beta_2)} + r_1 r_2 r_4 e^{-2i\beta_2} + r_1 r_2 r_5 e^{-2i(\beta_2+\beta_4)} + r_1 r_3 r_4 + r_1 r_3 r_5 e^{-2i\beta_4} +$

44  $r_1 r_4 r_5 e^{-2i(\beta_4-\beta_3)} + r_1 r_2 r_3 r_4 r_5 e^{-2i(\beta_4-\beta_3+\beta_2)} + r_2 r_3 r_4 e^{-2i\beta_1} + r_2 r_3 r_5 e^{-2i(\beta_1+\beta_4)} +$

45  $r_2 r_4 r_5 e^{-2i(\beta_1-\beta_3+\beta_4)} + r_3 r_4 r_5 e^{-2i(\beta_1+\beta_2-\beta_3+\beta_4)}$ (S6)

46  $r_b = e^{2i\beta_3} + r_1 r_2 e^{2i(\beta_3-\beta_1)} + r_1 r_3 e^{2i(\beta_3-\beta_2-\beta_1)} + r_1 r_4 e^{-2i(\beta_1+\beta_2)} + r_1 r_5 e^{-2i(\beta_1+\beta_2+\beta_4)} +$

47  $r_2 r_3 e^{2i(\beta_3-\beta_2)} + r_2 r_4 e^{-2i\beta_2} + r_2 r_5 e^{-2i(\beta_2+\beta_4)} + r_3 r_4 + r_3 r_5 e^{-2i\beta_4} + r_4 r_5 e^{-2i(\beta_4-\beta_3)} +$

48  $r_2 r_3 r_4 r_5 e^{-2i(\beta_4-\beta_3+\beta_2)} + r_1 r_2 r_3 r_4 e^{-2i\beta_1} + r_1 r_2 r_3 r_5 e^{-2i(\beta_1+\beta_4)} + r_1 r_2 r_4 r_5 e^{-2i(\beta_1-\beta_3+\beta_4)} +$

49  $r_1 r_3 r_4 r_5 e^{-2i(\beta_1+\beta_2-\beta_3+\beta_4)}$ (S7)

where $r_i=(n_{i-1}-n_i)/(n_{i-1}+n_i)$ is the Fresnel reflection coefficients at the interface from medium $i$-1 to $i$; the indices are assigned as air (0), MoS$_2$ (1), Au (2), Ti (3), SiO$_2$ (4), and Si (5). $n_i$ is the complex refractive index of the $i$th layer with $n_0 = 1$. $\beta_i = 2\pi n_i d_i/\lambda$ is the phase factor, representing the phase differences through the whole medium $i$, where $d_i$ is the thickness of medium $i$.